\documentclass[aps,prb,preprint,showpacs,preprintnumbers,amsmath,amssymb,floatfix]{revtex4} 

\usepackage{graphicx}
\usepackage{dcolumn}
\usepackage{bm}

\newcommand{\f}[2]{\ensuremath{\frac{\displaystyle{#1}}{\displaystyle{#2}}}} 
\newcommand{\lr}[1]{\langle{#1}\rangle} 

\begin{document}
\title{Phonon Band Structure and Thermal Transport Correlation in a Layered Diatomic Crystal}

\author{A. J. H. McGaughey\footnote{Current affiliation:
Department of Mechanical Engineering, Carnegie Mellon University,
Pittsburgh, PA 15213}}   \email{mcgaughey@cmu.edu}
\author{M. I. Hussein\footnote{Current affiliation: Department of Engineering,
University of Cambridge, Trumpington Street, CB2 1PZ, U.K.}}
\affiliation{Department of Mechanical Engineering, University of
Michigan, Ann Arbor, MI 48109-2125}

\author{E. S. Landry}
\affiliation{Department of Mechanical Engineering, Carnegie Mellon
University, Pittsburgh, PA 15213-3890}

\author{M. Kaviany}
\author{G. M. Hulbert}
\affiliation{Department of Mechanical Engineering, University of
Michigan, Ann Arbor, MI 48109-2125 }

\date{\today}

\begin{abstract}

To elucidate the relationship between a crystal's structure, its
thermal conductivity, and its phonon dispersion characteristics, an
analysis is conducted on layered diatomic Lennard-Jones crystals
with various mass ratios. Lattice dynamics theory and molecular
dynamics simulations are used to predict the phonon dispersion
curves and the thermal conductivity. The layered structure generates
directionally dependent thermal conductivities lower than those
predicted by density trends alone. The dispersion characteristics
are quantified using a set of novel band diagram metrics, which are
used to assess the contributions of acoustic phonons and optical
phonons to the thermal conductivity. The thermal conductivity
increases as the extent of the acoustic modes increases, and
decreases as the extent of the stop bands increases. The sensitivity
of the thermal conductivity to the band diagram metrics is highest
at low temperatures, where there is less anharmonic scattering,
indicating that dispersion plays a more prominent role in thermal
transport in that regime. We propose that the dispersion metrics (i)
provide an indirect measure of the relative contributions of
dispersion and anharmonic scattering to the thermal transport, and
(ii) uncouple the standard thermal conductivity structure-property
relation to that of structure-dispersion and dispersion-property
relations, providing opportunities for better understanding of the
underlying physical mechanisms and a potential tool for material
design.

\end{abstract}

\pacs{63.20.Dj}

\maketitle

\section{\label{S-intro}Introduction}

Thermal transport in a dielectric crystal is governed by phonon
dispersion and phonon scattering.\cite{scatter} The majority of
theoretical studies of thermal transport in dielectrics deal with
phonon dispersion at a qualitative level. A common treatment is to
assume that the contribution of optical phonons to the thermal
conductivity is negligible because the associated dispersion
branches are often flat, implying low phonon group velocities.
Theories that quantitatively relate dispersion characteristics to
bulk thermal transport properties are limited. One example is the
use of phonon dispersion curves to determine the phonon group and
phase velocities required in the single mode relaxation time
formulation of the Boltzmann transport equation
(BTE).\cite{callaway1959,holland1963} Even in such a case, the
dispersion has traditionally been greatly simplified. The importance
of accurately and completely incorporating dispersion into this BTE
formulation has recently been investigated for bulk
materials\cite{chung2004,mcgaughey2004c} and for
nanostructures.\cite{mazumder2001,mingo2003}

Dong et al.\cite{dong2001} report evidence of the important role
that phonon dispersion plays in thermal transport in their study of
germanium clathrates using molecular dynamics (MD) simulations. In
their Fig$.$ 1, they show phonon dispersion curves for a diamond
structure, a clathrate cage, and the same cage structure but filled
with weakly bonded guest strontium atoms that behave as ``rattlers."
While the range of frequencies accessed by the vibrational modes in
these three structures is comparable, the dispersion characteristics
are quite different. The large unit cell of the clathrate cage
significantly reduces the frequency range of the acoustic phonons,
the carriers generally assumed to be most responsible for thermal
transport. There is an accompanying factor of ten reduction in the
thermal conductivity. In the filled cage, the encapsulated guest
atoms have a natural frequency that cuts directly through the middle
of what would be the acoustic phonon branches, and the value of the
thermal conductivity is reduced by a further factor of ten.
Experimental studies on filled cage-like structures have found
similar results, i.e., that rattler atoms can reduce the thermal
conductivity.\cite{nolas1998,cohn1999}

Considering phonon dispersion is also important in studying thermal
transport in superlattices
(SLs).\cite{broido2004,simkin2000,chen2005,chen2004} Using an
inelastic phonon Boltzmann approach to model anharmonic three-phonon
scattering processes, Broido and Reinecke\cite{broido2004} studied
the thermal conductivity of a model two-mass SL with a diamond
structure, and how it depends on the mass ratio and the layer
thickness. As the mass ratio is increased, the dispersion curves
flatten, which tends to lower the thermal conductivity. At the same
time, the increase in mass ratio reduces the cross section for
Umklapp scattering, which tends to increase the thermal
conductivity. The relative importance of these two mechanisms is
found to be a function of the layer thickness. Using a kinetic
theory model, Simkin and Mahan\cite{simkin2000} found that as the
layer thickness in a model SL is reduced, the thermal conductivity
decreases due to an increase in ballistic scattering stemming from
the rise in interface density. It was shown in the same study that
as the layer thickness is further reduced to values sufficiently
smaller than the phonon mean free path, the thermal conductivity
increases. This transition, which predicts a minimum SL thermal
conductivity, was attributed to a shift from phonon transport best
described by a particle theory to one that follows a wave theory.
This minimum SL thermal conductivity has also been measured in
experiments\cite{venkat2000,chakraborty2003,caylor2005} and
predicted in MD simulations.\cite{chen2005} The need for a wave
treatment of phonons indicates that interference mechanisms affect
the phonon transport.

\begin{figure}
\includegraphics{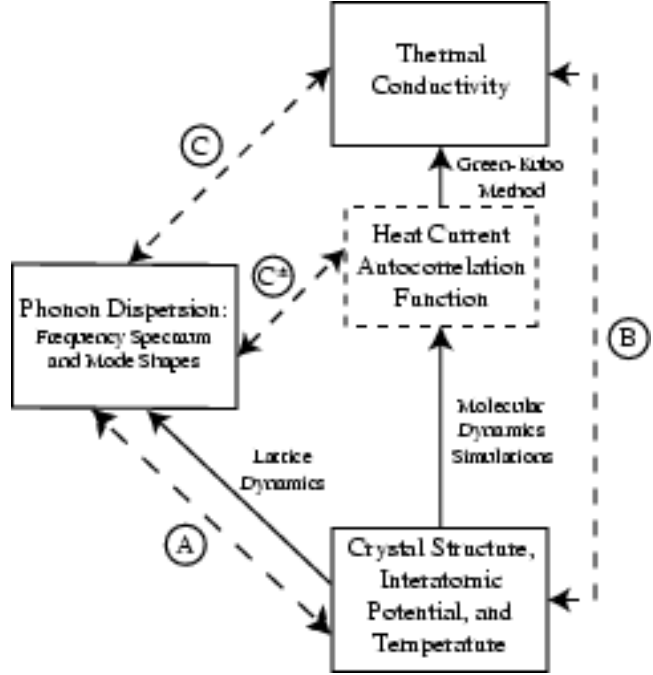}
\caption{\label{F-3way} Schematic of the relationships between
atomic structure, dispersion characteristics, and thermal
conductivity. Shown are both the tools available for moving between
these blocks (solid lines), and the links we are seeking to
establish (dashed lines). Link A is discussed in Section
\ref{S-dispersion}, link B in Section \ref{SS-pred}, link C$^*$ in
Section \ref{SS-disp-hcacf}, and link C in Section \ref{SS-disp-k}.
In this investigation, we restrict our discussion of the lattice
dynamics to the phonon mode frequencies.}
\end{figure}

Apart from the above mentioned efforts, few investigations have
attempted to rigorously establish a connection between dispersion
characteristics, which include the size and location of frequency
bands (and stop bands), and thermal transport properties. In this
work we explore the three-way relationship in a crystal between: (i)
the unit cell structure, (ii) the associated dispersion
characteristics, and (iii) the bulk thermal transport behavior using
lattice dynamics calculations and MD simulations, as shown in Fig$.$
\ref{F-3way}. The dispersion characteristics are quantified using a
new set of frequency band diagram metrics. As a starting point, we
narrow our attention to a diatomic Lennard-Jones (LJ) crystal that
corresponds to a monolayer SL. The atomic species are only
differentiated by their masses. By modeling a simple system,
phenomena can be observed that might not be discernable in more
complex structures. The overall theme of the investigation, however,
is intended for dielectric crystals in general and the analysis
tools developed are not limited to SLs.

The insights gained in this study could lead to the development of a
systematic technique for the atomic-level design of materials with
desired thermal transport properties. This capability could
facilitate the introduction of novel, yet realizable, materials with
very high, or low, thermal conductivities. Examples of applications
include thermoelectric materials with high figure-of-merit,
microelectronic devices enjoying enhanced cooling characteristics,
and efficient thermal insulators for chemical processing.

We begin by presenting the diatomic crystal structure and basic
information pertaining to the MD simulations. Phonon dispersion
relations are then determined using lattice dynamics calculations
and analyzed for different mass ratios at various temperatures. The
band diagram metrics are introduced and discussed. We then use MD
simulations and the Green-Kubo (GK) method to predict the thermal
conductivities of these structures. Discussion is presented with
regards to the magnitude of the thermal conductivity, its
directional and temperature dependencies, and relation to the unit
cell. We then explore the relationship between the thermal
conductivity and the associated dispersion band structure.

\section{\label{S-structure} Crystal structure and molecular dynamics simulations}

We perform our study of the relationship between atomic structure,
phonon band structure, and thermal transport by considering model
systems described by the LJ potential,
\begin{equation}
\phi_{ij}(r_{ij}) = 4\epsilon_{\mathrm{LJ}}
\left[\left(\frac{\sigma_{\mathrm{LJ}}}{r_{ij}}\right)^{12} -
\left(\frac{\sigma_{\mathrm{LJ}}}{r_{ij}}\right)^6\right].
\label{E-LJ}
\end{equation}
Here, $\phi_{ij}$ is the potential energy associated with a pair of
particles ($i$ and $j$) separated by a distance $r_{ij}$, the
potential well depth is $\epsilon_{\mathrm{LJ}}$, and the
equilibrium pair separation is $2^{1/6} \sigma_{\mathrm{LJ}}$. The
LJ potential is commonly used in investigations of thermal transport
in bulk and composite
systems.\cite{lukes2000,abramson2002,chantrenne2004,lukes2004,chen2004,mcgaughey2004a,mcgaughey2004c,mcgaughey2005,chen2005}
Dimensional quantities have been scaled for argon, for which
$\epsilon_{\mathrm{LJ}}$ and $\sigma_{\mathrm{LJ}}$ have values of
$1.67 \times 10^{-21}$ J and $3.4 \times 10^{-10}$
m.\cite{ashcroft1976} The argon mass scale, $m_{\mathrm{LJ}}$, is
$6.63 \times 10^{-26}$ kg (the mass of one atom). Quantities
reported in dimensionless LJ units are indicated by the superscript
*.

\begin{figure}
\includegraphics{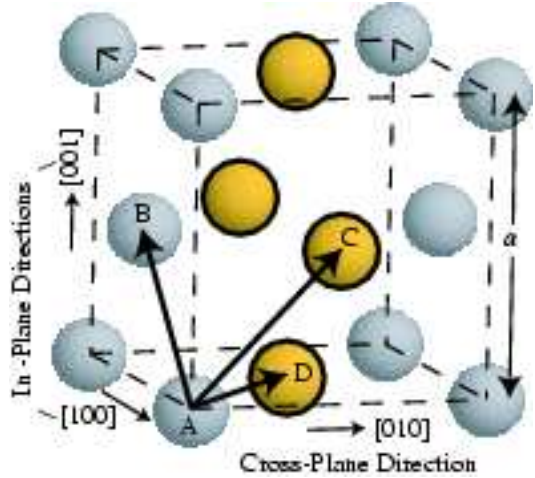}
\caption{\label{F-structure} The four-atom unit-cell of the layered
diatomic crystal. If the four atoms are equivalent, the crystal
structure is face centered cubic. If A and B have a different mass
than C and D, the crystal structure is simple cubic. The arrangement
chosen leads to planes with a thickness of one atomic layer stacked
in the [010] (cross-plane) direction. (Color online)}
\end{figure}

The structure we consider is a simple-cubic crystal with a four-atom
basis (unit cell) and lattice parameter $a$, as shown in Fig$.$
\ref{F-structure}. If the four atoms A-D are identical, the
conventional unit-cell representation of the face-centered cubic
(fcc) crystal structure is obtained. By taking two of the atoms in
the basis (A, B) to have one mass, $m_1$, and the other two (C, D)
to have a different mass, $m_2$, a crystal with alternating layers
of atoms in the [010] direction [which we will refer to as the
cross-plane (CP) direction] is created. We will refer to the [100]
and [001] directions as the in-plane (IP) directions. This structure
is a monolayer SL. The value of $m_2$ is fixed at $m_{\mathrm{LJ}}$,
and the mass ratio $R_m$ is defined as
\begin{eqnarray}
R_m \equiv \f{m_1}{m_2}. \label{E-massratio}
\end{eqnarray}
The symbol $m^*$ will be used to denote the dimensionless mass of
the atoms in a monatomic system. It has a value of unity, unless
noted. By varying only the atomic masses, the LJ potential can be
applied to this system without any modification. We primarily
consider $R_m$ values of 1 (the monatomic unit cell), 2, 5, and 10.
We note that this variation in atomic masses is in the same spirit
as the work of Broido and Reinecke\cite{broido2004} and Che et
al.,\cite{che2000} who considered two-mass diamond systems. Their
approach was different, though, in that they varied the mass ratio
while keeping the density constant.

The MD simulations are performed in dimensionless LJ units. All
results can be scaled to dimensional values using the parameters
$\epsilon_{\mathrm{LJ}}$, $\sigma_{\mathrm{LJ}}$, and
$m_{\mathrm{LJ}}$, and the Boltzmann constant, $k_{\mathrm B}$,
which is part of the temperature scale, $\epsilon_{\mathrm{LJ}} /
k_{\mathrm B}$. From dimensional analysis, it can be shown that the
thermal conductivity of the $1/R_m$ system, $k_{1/R_m}$, is related
to that of the $R_m$ system, $k_{R_m}$, through
\begin{eqnarray}
k_{1/R_m} = k_{R_m} R_m^{1/2}.
\end{eqnarray}
Thus, one dimensionless simulation can be used to generate data for
both the $R_m$ system and the $1/R_m$ system. Where helpful, data
corresponding to mass ratios of 0.5, 0.2, and 0.1 will be presented.

The purpose of the MD simulations is (i) to obtain the zero-pressure
lattice parameter as a function of temperature (used in later
simulations and for lattice dynamics calculations), and (ii) to
predict the thermal conductivity. Because the functional form of the
LJ potential is kept the same (we consider only different masses),
the zero-pressure unit cell parameters remain unchanged from those
for a single mass system.\cite{mcgaughey2004c}

All reported MD results are generated from simulations run in the
$NVE$ (constant mass, volume and energy) ensemble at zero pressure
with a time step of 4.285 fs. The simulation cell contains 512 atoms
(eight unit cells in the cross-plane direction, and four in the
in-plane directions) and periodic boundary conditions are imposed in
all three directions. This system size was found to be sufficient to
obtain computational domain-independent thermal conductivities. The
MD procedures have been described in detail
elsewhere.\cite{mcgaughey2004c,mcgaughey2004a,mcgaugheythesis} We
consider temperatures, $T$, between 10 K and 80 K in 10 K
increments. The melting temperature of the monatomic system is 87 K.
The data presented in the thermal conductivity analysis for a given
parameter set ($R_m$, $T$, and the crystallographic direction) are
obtained by averaging over five independent simulations so as to get
a representative sampling of phase space. In each of the five
simulations, information is obtained over one million time steps,
after an equilibration period of five hundred thousand time steps.

\section{\label{S-dispersion}Phonon dispersion}

The frequency (phonon) space characteristics of a solid phase can be
determined with lattice dynamics calculations,\cite{dove1993} in
which the real space coordinates (the positions) are transformed
into the normal mode coordinates (the phonon modes). Each normal
mode has a frequency, $\omega$, wave vector, $\pmb{\kappa}$, and
polarization vector, $\mathbf{e}$ (which describes the mode shape).
The available wave vectors are obtained from the crystal structure,
and the frequencies and polarization vectors are found by solving
the eigenvalue equation\cite{dove1993}
\begin{eqnarray}
\omega^2(\pmb{\kappa},\nu) \mathbf{e}(\pmb{\kappa},\nu) =
\mathbf{D}(\pmb{\kappa}) \cdot \mathbf{e}(\pmb{\kappa},\nu),
\label{E-eigenvalue}
\end{eqnarray}
where $\mathbf{D}(\pmb{\kappa})$ is the dynamical matrix for the
unit cell and the parameter $\nu$ identifies the polarization for a
given wave vector. The lattice dynamics methods used are reviewed in
the Appendix.

The predictions of harmonic lattice dynamics calculations from Eq$.$
(\ref{E-eigenvalue}) at zero-temperature are exact, and can be found
using the equilibrium atomic positions and the interatomic
potential. As the temperature rises, the system will move away from
the zero-temperature minimum of the potential energy surface. The
lattice dynamics are affected in two ways. First, the solid will
either expand or contract. This effect can be accounted for by using
the finite-temperature lattice constant in the harmonic lattice
dynamics formulation (the quasi-harmonic
approximation).\cite{dove1993} Second, the motions of the atoms will
lead to anharmonic interactions. These effects cannot be easily
accounted for in the formal lattice dynamics theory due to the
difficulty in incorporating third order (and higher) derivatives of
the interatomic potential. In an MD simulation, it is possible, with
considerable effort, to capture the true anharmonic lattice
dynamics.\cite{mcgaughey2004c} For the purposes of this
investigation, however, we restrict the analysis to the
quasi-harmonic formulation. Further discussion of this choice will
follow where appropriate.

\begin{figure}
\includegraphics{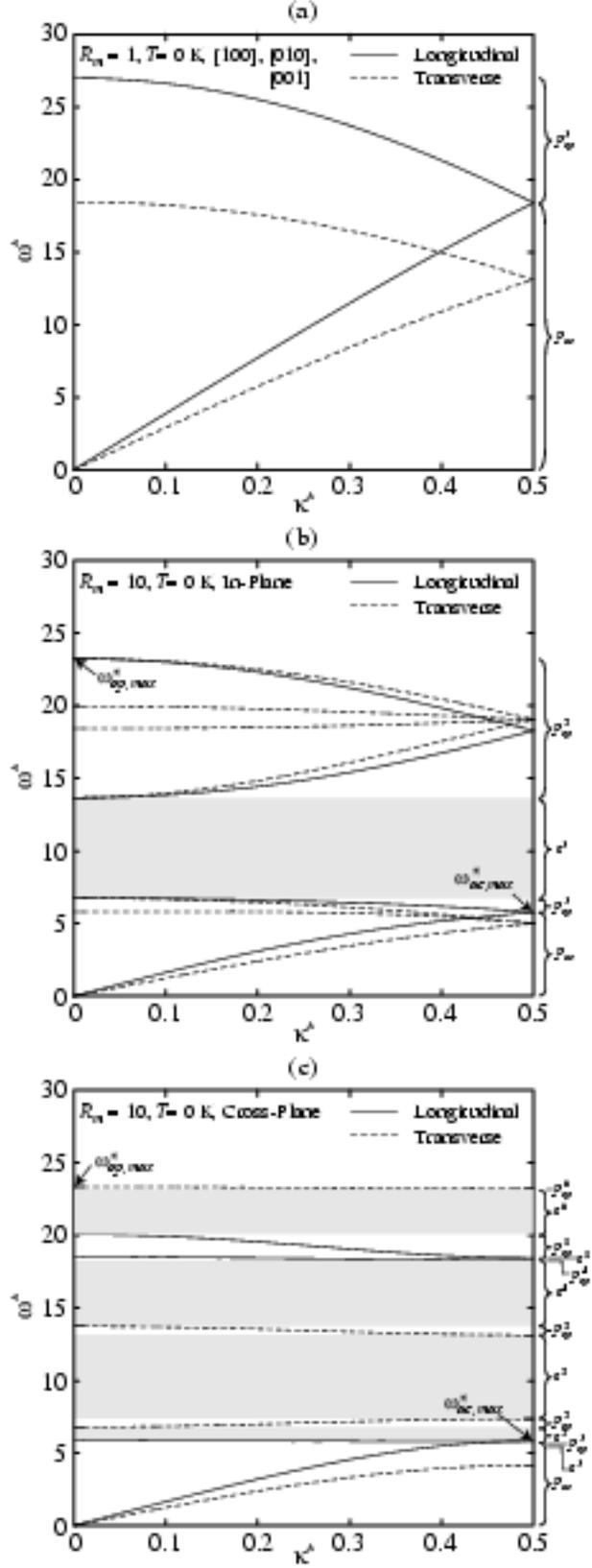}
\caption{\label{F-disp} (a) Phonon dispersion curves for the
monatomic crystal. The dispersion characteristics are the same in
the [100], [010], and [001] directions. All branches correspond to
acoustic phonons, and have been folded over at $\kappa^* = 0.5$.
Phonon dispersion curves for $R_m = 10$: (b) in-plane and (c)
cross-plane directions. The maximum acoustic and optical frequencies
are noted, and the stop bands are shaded.}
\end{figure}

We now consider the relationship between the mass ratio $R_m$ and
the dispersion characteristics (A in Fig$.$ 1) by plotting a set of
dispersion curves in Figs$.$ \ref{F-disp}(a), \ref{F-disp}(b), and
\ref{F-disp}(c). The plots in Fig$.$ \ref{F-disp} are of
dimensionless frequency, $\omega^*=\omega / (\epsilon_{\mathrm{LJ}}
/ m_{\mathrm{LJ}} \sigma_{\mathrm{LJ}}^2)^{1/2} $, versus
dimensionless wave number, defined as $\kappa^* = \kappa / (2 \pi /
a)$. The zero-temperature phonon dispersion curves for the monatomic
($R_m=1$) unit cell in the [100] direction are shown in Fig$.$
\ref{F-disp}(a).  As this system has cubic isotropy, the dispersion
characteristics are the same for the [010] and [001] directions. For
the monatomic crystal, the true Brillouin zone extends to a
$\kappa^*$ value of unity in the [100] direction and is a truncated
octahedron. Here, to be consistent with the plots for the layered
structure (where the Brillouin zone extends to a $\kappa^*$ value of
0.5 and is cubic), the branches have been folded over at their
midpoints.

The zero-temperature dispersion curves for the in-plane and
cross-plane directions for the $R_m$ = 10 unit cell are shown in
Figs$.$ \ref{F-disp}(b) and \ref{F-disp}(c). The anisotropy of the
crystal structure is reflected in the directional dependence of the
plotted dispersion curves. At non-unity mass ratios, stop bands
(band gaps) in the frequency spectra are evident, and grow as $R_m$
is increased. A stop band in the in-plane direction opens when the
mass ratio is 1.17 and in the cross-plane direction when the mass
ratio is 1.62. The stop bands in Figs$.$ \ref{F-disp}(b) and
\ref{F-disp}(c) are shaded gray.

The dispersion curves plotted correspond to a crystal of infinite
size, and are thus continuous. Due to the finite size of the
simulation cell, only the modes at $\kappa^*$ values of 0, 0.25 and
0.5 will be present in the MD simulations for the in-plane
directions, and at $\kappa^*$ values of 0, 0.125, 0.25, 0.375, and
0.5 for the cross-plane direction.

\begin{figure}
\includegraphics{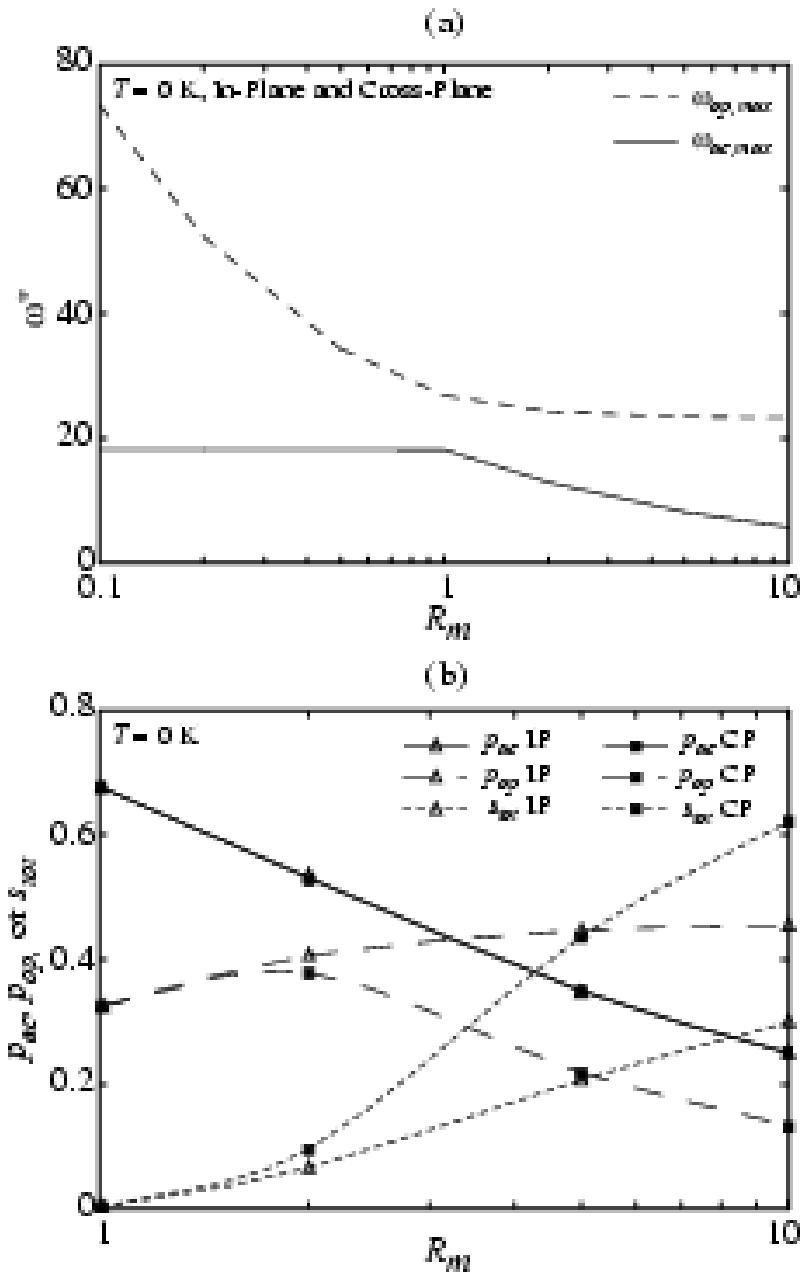}
\caption{\label{F-disp-gen} (a) Maximum frequencies of the acoustic
and optical branches plotted against the mass ratio $R_m$ at zero
temperature. The data in the in-plane and cross-plane directions are
indistinguishable. (b) Breakdown of the frequency spectrum into the
acoustic and optical pass bands ($p_{ac}$ and $p_{op}$) and the stop
bands, $s_{all}$. The $p_{ac}$ curves for the two directions are
indistinguishable. The data have been fit with smoothed curves to
highlight the trends.}
\end{figure}

The maximum frequencies of the acoustic and optical branches at
zero-temperature are plotted in Fig$.$ \ref{F-disp-gen}(a) for $R_m$
values between 0.1 and 10. The data for the in-plane and cross-plane
directions are indistinguishable. Maximum frequency values for the
monatomic system are taken at $\kappa^*$ values of 0.5 and 1 for the
acoustic and optical branches. As the mass ratio is increased from
unity, the maximum optical frequency decreases slightly and the
maximum acoustic frequency decreases. As $R_m$ is reduced below
unity, the maximum optical frequency increases and the maximum
acoustic frequency is essentially constant. The maximum optical
frequency increases for small $R_m$ because the lighter mass can
oscillate at higher frequencies. These trends are qualitatively
consistent with those predicted for a one-dimensional diatomic
system.\cite{kittel1996}

The LJ system expands as its temperature is raised, causing the
phonon frequencies to decrease. At a temperature of 80 K, the
quasi-harmonic dispersion calculations for the monatomic system give
a maximum dimensionless frequency, $\omega_{op,max}^*$, of 16.6,
compared to the zero-temperature maximum of 26.9. The temperature
dependence of the maximum frequency is close to linear, and a
similar percentage decrease is found for the non-unity mass ratios.
The full anharmonic dispersion analysis for the monatomic system at
a temperature of 80 K gives a maximum dimensionless frequency of
21.2.\cite{mcgaughey2004c} The discrepancy between the
quasi-harmonic and anharmonic dispersion curves increases with
increasing temperature and wave vector magnitude. For the majority
of the data considered here, the expected difference is less than
10\%, which we take to be satisfactory in lieu of calculating the
full anharmonic dispersion characteristics.

In later analysis, we will investigate the relationship between
thermal conductivity and the dispersion curves. To facilitate this
analysis, we introduce three metrics for the dispersion curves:
$p_{ac}$, $p_{op}$, and $s_{all}$. The idea of parameterizing the
frequency band structure has previously been applied in the area of
elastodynamics.\cite{hussein2006} The first of the metrics,
$p_{ac}$, is defined as the fraction of the frequency spectrum taken
up by the acoustic branches, i.e., $p_{ac} = \omega_{ac,max} /
\omega_{op,max}$. This is the acoustic-phonon pass band. Similarly,
$p_{op}$ represents the portion of the spectrum taken up by the
optical branches [the optical-phonon pass band(s)]. We calculate
$p_{op}$ as $\sum_i p_{op}^i$, where the summation is over the
optical phonon bands, as shown in Figs$.$ \ref{F-disp}(b) and
\ref{F-disp}(c). The rest of the spectrum is made up of stop bands,
whose contribution is given by $s_{all} = \sum_i s^i$, as shown in
Figs$.$ \ref{F-disp}(b) and \ref{F-disp}(c). Thus,
\begin{eqnarray}
p_{ac} + p_{op} + s_{all} = 1.
\end{eqnarray}
These three quantities are plotted as a function of the mass ratio
for both the in-plane and cross-plane directions at zero temperature
in Fig$.$ \ref{F-disp-gen}(b). As the mass ratio increases, the
acoustic phonons take up less of the spectrum (the results are
essentially identical in the two directions), while the extent of
the stop bands increases (more so in the cross-plane direction). The
optical-phonon extent initially increases in both directions. This
increase continues for the in-plane direction, but a decrease is
found for the cross-plane direction, consistent with its larger stop
band extent. The temperature dependence of $p_{ac}$, $p_{op}$, and
$s_{all}$ is small ($\sim$ 1\% change between temperatures of 0 K
and 80 K).

\section{\label{S-k} Thermal conductivity}

In this section we address the relationship between atomic structure
and thermal conductivity (B in Fig$.$ 1), with minimal consideration
of the phonon dispersion.

\subsection{\label{SS-GK} Green-Kubo method}

The net flow of heat in the MD system, given by the heat current
vector $\mathbf{S}$, fluctuates about zero at equilibrium. In the GK
method, the thermal conductivity is related to how long it takes for
these fluctuations to dissipate. For the layered crystal under
consideration, the thermal conductivity will be anisotropic
(different in the in-plane and cross-plane directions) and for a
direction $l$ will given by \cite{mcquarrie2000}
\begin{equation}
k_l = \frac{1}{k_{\mathrm B} V T^2} \int_0^\infty \lr{S_l(t) S_l(0)}
dt, \label{E-kGK}
\end{equation}
where $t$ is time and $\lr{S_l(t) S_l(0)}$ is the heat current
autocorrelation function (HCACF) for the direction $l$. The heat
current vector for a pair potential is given by \cite{mcquarrie2000}
\begin{equation}
\mathbf{S} = \frac{d}{dt}\sum_i E_i \mathbf{r}_i =  \sum_i E_i
\mathbf{v}_i + \frac{1}{2} \sum_{i} \sum_{j \ne i} (\mathbf{F}_{ij}
\cdot \mathbf{v}_i)\mathbf{r}_{ij},\label{E-q}
\end{equation}
where $E_i$, $\mathbf{r}_i$, and $\mathbf{v}_i$ are the energy,
position vector, and velocity vector of particle $i$, and
$\mathbf{r}_{ij}$ and $\mathbf{F}_{ij}$ are the inter-particle
separation vector and force vector between particles $i$ and $j$.

The most significant challenge in the implementation of the GK
method is the specification of the integral in Eq$.$ (\ref{E-kGK}),
which may not converge due to noise in the data. Such an event may
be a result of not obtaining a proper sampling of the system's phase
space, even when averaging over a number of long, independent
simulations, as we have done here. Thus, it is not always possible
to directly specify the thermal conductivity. As such, we also
consider finding the thermal conductivity by fitting the HCACF to a
set of algebraic functions.

It has been shown \cite{mcgaughey2004a,mcgaughey2004b} that the
thermal conductivities of the monatomic LJ fcc crystal and a family
of complex silica structures can be decomposed into contributions
from short and long time-scale interactions by fitting the HCACF to
a function of the form
\begin{eqnarray}
\langle S_l(t)S_l(0)\rangle = A_{ac,sh,l} \exp(-t/\tau_{ac,sh,l})+
A_{ac,lg,l} \exp(-t/\tau_{ac,lg,l}) + \label{E-qdecomp}\\
\sum_j B_{op,j,l} \exp(-t/ \tau_{op,j,l}) \cos(\omega_{op,j,l} t).
\nonumber
\end{eqnarray}
Here, the subscripts $sh$ and $lg$ refer to short-range and
long-range, the $A$ and $B$ parameters are constants, and $\tau$
denotes a time constant. The summation in the optical phonon term
corresponds to the peaks in the Fourier transform of the
HCACF,\cite{mcgaughey2004b} where $\omega_{op,j,l}$ is the frequency
of the $j$th peak in the $l$th direction. This term is used when
appropriate (i.e., for crystals with more than one atom in the unit
cell). Substituting Eq$.$ (\ref{E-qdecomp}) into Eq$.$
(\ref{E-kGK}),
\begin{eqnarray}
k_l &=&  \frac{1}{k_{\mathrm B} V T^2}\left(A_{ac,sh,l}
\tau_{ac,sh,l} + A_{ac,lg,l} \tau_{ac,lg,l} +  \sum_j \f{B_{op,j,l}
\tau_{op,j,l}}{1+
\tau_{op,j,l}^2 \omega_{op,j,l}^2}\right) \label{E-kdecomp}\\
&\equiv&k_{ac,sh,l} + k_{ac,lg,l} + k_{op,l}. \nonumber
\end{eqnarray}
Further details on the decomposition can be found elsewhere.
\cite{mcgaughey2004a,mcgaughey2004b} In this investigation, a fit of
Eq$.$ (\ref{E-qdecomp}) was possible for all cases except the
cross-plane direction for (i) $R_m=5$ ($T \ge 60 K$) and (ii)
$R_m=10$ ($T \ge 40 K$). In these cases, the peak in the spectrum of
the HCACF broadens to the point where it is no longer well
represented by an exponentially decaying cosine function in the time
coordinate.\cite{fit} The agreement between the fit thermal
conductivity and that specified directly from the integral is
generally within 10\%. When available, the fit values are reported.

\subsection{\label{SS-HCACF} Heat current autocorrelation function}

\begin{figure}
\includegraphics{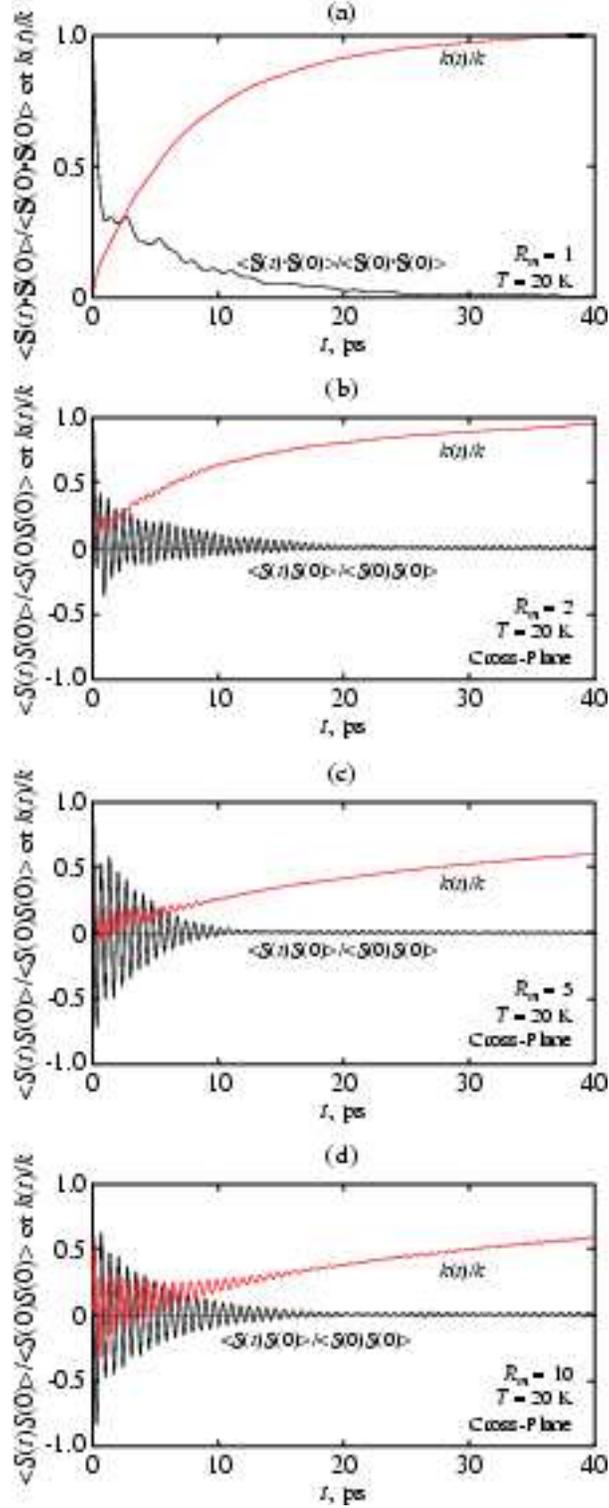}
\caption{\label{F-hcacf} The HCACF and its integral (whose converged
value is the thermal conductivity) for $R_m =$ (a) 1, (b) 2, (c) 5,
and (d) 10 at a temperature of 20 K. The data for the latter three
cases corresponds to the cross-plane direction. The HCACF is scaled
against the zero time value, while the thermal conductivity is
scaled against its converged long-time value. (Color online)}
\end{figure}

The time dependence of the HCACF (normalized by its zero-time
value), and its integral (the thermal conductivity, normalized by
its converged value), are plotted for four cases at a temperature of
20 K in Figs$.$ \ref{F-hcacf}(a), \ref{F-hcacf}(b),
\ref{F-hcacf}(c), and \ref{F-hcacf}(d).

The data in Fig$.$ \ref{F-hcacf}(a) correspond to the monatomic
system. The HCACF decays monotonically (small oscillations can be
attributed to the periodic boundary
conditions\cite{mcgaugheythesis}), and is well fit by the sum of two
decaying exponentials [the first two terms in Eq$.$
(\ref{E-qdecomp})]. Such behavior is found at all temperatures
considered for the monatomic system.

The data in Figs$.$ \ref{F-hcacf}(b), \ref{F-hcacf}(c), and
\ref{F-hcacf}(d) correspond to the cross-plane direction for $R_m$
values of 2, 5, and 10. Significant oscillations in the HCACF are
present, and grow as the mass ratio is increased. This is the effect
of the optical phonon modes.\cite{mcgaughey2004b}  The HCACF can be
fit by Eq$.$ (\ref{E-qdecomp}), with one term taken in the optical
summation (this is justified by the HCACF spectra shown later). The
integrals in Figs$.$ \ref{F-hcacf}(c) and \ref{F-hcacf}(d) converge
at a longer time than shown in the plot. This result indicates that
while the oscillations in the HCACF dominate its magnitude, the
contribution of the $k_{ac,lg}$ term to the thermal conductivity is
significant, even if it cannot be visually resolved. The plots in
Fig$.$ \ref{F-hcacf} show the effect of changing the mass ratio at a
fixed temperature. For a fixed mass ratio, increasing the
temperature leads to a HCACF that decays faster. Fewer periods are
captured in the decay, which makes it harder to extract the
oscillation frequency. Results for the in-plane direction are
qualitatively similar.

\begin{figure}
\includegraphics{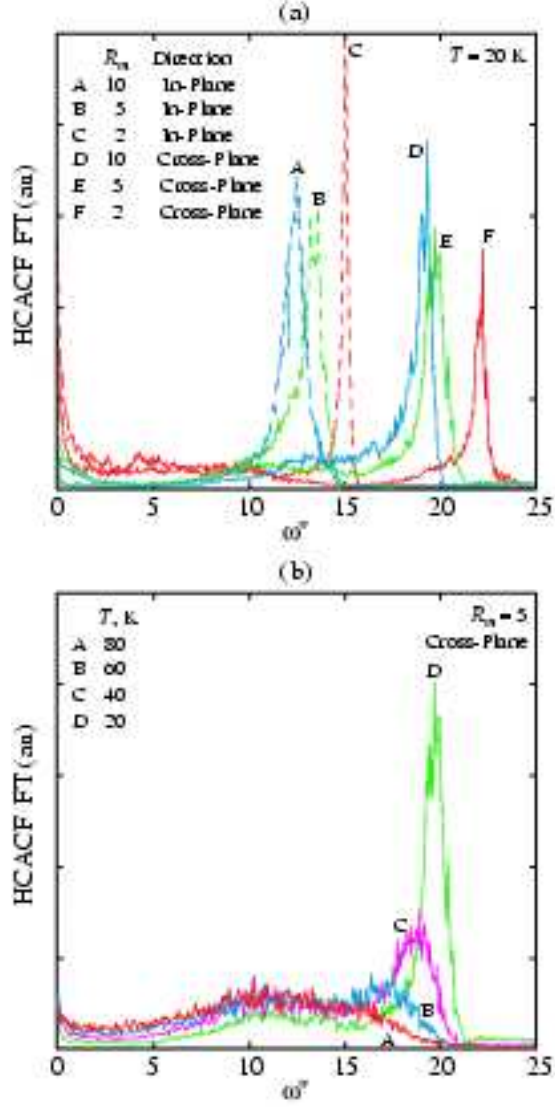}
\caption{\label{F-whcacf} HCACF frequency spectrum for (a) $T = 20$
K, mass ratios of 2, 5, and 10 in both the in-plane and cross-plane
directions and (b) mass ratio of 5, cross-plane direction, for
temperatures of 20 K, 40 K, 60 K, and 80 K. (Color online)}
\end{figure}

In Fig$.$ \ref{F-whcacf}(a), the Fourier transform (FT) of the
normalized HCACF is plotted for mass ratios of 2, 5, and 10 in the
in-plane and cross-plane directions at a temperature of 20 K. Each
of the spectra have a non-zero zero-frequency intercept, and a
strong, single peak. This is in contrast to the FT of the monatomic
system, which decays monotonically from the zero-frequency
value.\cite{mcgaughey2004a} The peaks fall within the range of the
optical phonons, and the peak frequency is best found through the
HCACF decomposition. As the mass ratio increases, the location of
the peak frequency in each of the directions decreases, which is
consistent with the trends in the dispersion curves plotted in
Fig$.$ \ref{F-disp-gen}(a).

In Fig$.$ \ref{F-whcacf}(b), the FT of the normalized HCACF for a
mass ratio of 5 in the cross-plane direction is plotted at
temperatures of 20 K, 40 K, 60 K, and 80 K. As the temperature
increases, the peak frequency decreases, consistent with the
discussion of the effects of temperature on phonon frequencies in
Section \ref{S-dispersion}. The peak also broadens considerably and
its magnitude decreases; the behavior is better described as
broad-band excitation. It is this behavior that leads to the failure
of the fitting procedure under certain conditions. This broadening
is not as significant for a mass ratio of 2. The origin of the peaks
will be further discussed in Section \ref{SS-disp-hcacf}.

\subsection{\label{SS-pred} Predictions}

\begin{figure}
\includegraphics{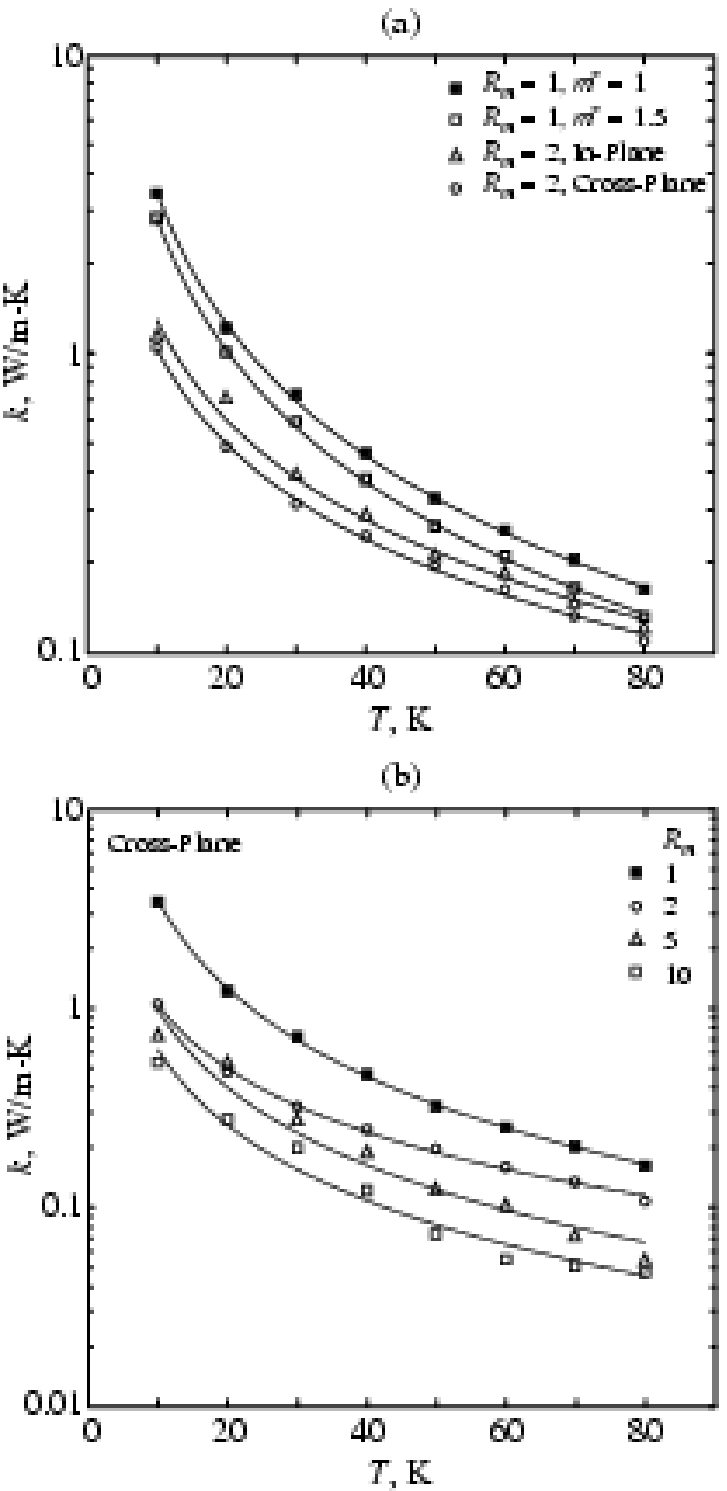}
\caption{\label{F-kT} (a) Thermal conductivity plotted as a function
of temperature for mass ratios of 1 and 2. (b) Thermal conductivity
in the cross-plane direction plotted as a function of temperature
for mass ratios of 1, 2, 5, and 10. The data in both (a) and (b) are
fit with power-law functions to guide the eye. }
\end{figure}

The thermal conductivities predicted for the $R_m = 2$ system are
plotted in Fig$.$ \ref{F-kT}(a) along with the data for the
monatomic unit cell. Also included in the plot are the predictions
for a monatomic cell with all atoms having a mass of 1.5 (i.e., $R_m
= 1$, $m^*=1.5$, giving the same density as the $R_m=2$ system).
These results are obtained by scaling the data from the ($R_m = 1$,
$m^*=1$) system. While increasing the density lowers the thermal
conductivity, the presence of the different masses has a more
pronounced effect for both the in-plane and cross-plane directions,
with the former being lower than the latter. The thermal
conductivities for all cases considered in the cross-plane direction
are plotted in Fig$.$ \ref{F-kT}(b). The data in Figs$.$
\ref{F-kT}(a) and \ref{F-kT}(b) are fit with power law functions to
guide the eye. Due to the classical nature of the simulations, all
the curves will go to infinity at zero temperature (unlike
experimental data, which peaks and then goes to zero at zero
temperature).

\begin{figure}
\includegraphics{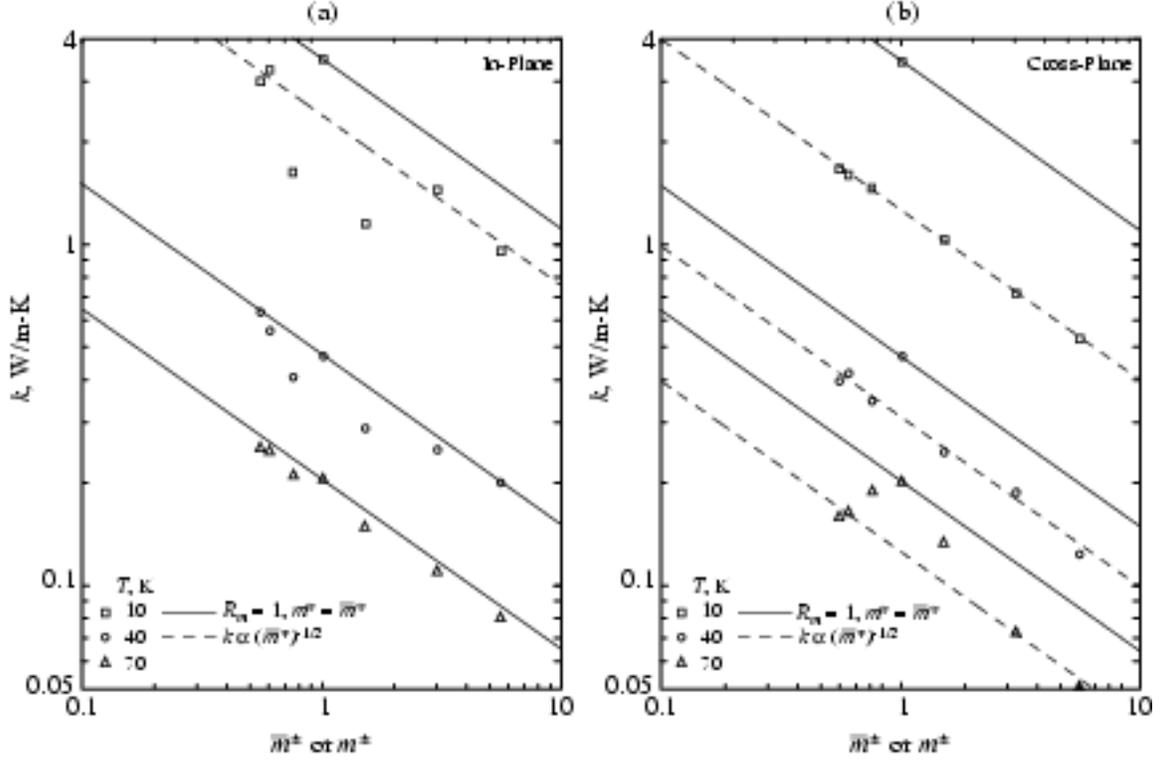}
\caption{\label{F-km} Thermal conductivity at temperatures of 10 K,
40 K, and 70 K in the (a) in-plane and (b) cross-plane directions
plotted as a function of the average atomic mass in the system. Also
plotted is the thermal conductivity of the equivalent monatomic
systems where each atom has a mass $\bar{m}^*$ (solid line). The
dashed lines correspond to $k \propto (\bar{m}^*)^{-1/2}$, and are
fit to the data by visual inspection.}
\end{figure}

By considering the LJ thermal conductivity scale,
$k_{\mathrm{B}}/\sigma_{\mathrm{LJ}}^2 (\epsilon_{\mathrm{LJ}} /
m_{\mathrm{LJ}})^{1/2}$, it can be deduced that the thermal
conductivity of a monatomic system where each atom has mass $m^*$ is
proportional to $(m^*)^{-1/2}$. We will denote the $k \propto
(m^*)^{-1/2}$ behavior as the density trend. In Figs$.$
\ref{F-km}(a) and \ref{F-km}(b), the thermal conductivities for mass
ratios between 0.1 and 10 in both directions at temperatures of 10
K, 40 K, and 70 K are plotted as a function of the average
dimensionless atomic mass in the unit cell, $\bar{m}^*$, given by
\begin{eqnarray}
\bar{m}^* = \f{m_1^*+ m_2^*}{2}, \label{E-mbarstar}
\end{eqnarray}
which is proportional to the system density. Also plotted for each
data series in Figs$.$ \ref{F-km}(a) and \ref{F-km}(b) is the
thermal conductivity that would exist in an equivalent monatomic
crystal where all atoms have mass $\bar{m}^*$ (i.e., $m^* =
\bar{m}^*$). These are the solid lines in the plots, obtained by
scaling the ($R_m = 1$, $m^*=1$) data point. The dashed lines in
Figs$.$ \ref{F-km}(a) and \ref{F-km}(b) correspond to $k \propto
(m^*)^{-1/2}$, and have been placed by eye to highlight trends. From
these two plots, the relationships between the effects of
temperature (i.e., anharmonicity and scattering), density, and the
mass ratio become apparent.

In the in-plane direction [Fig$.$ \ref{F-km}(a)], the thermal
conductivity data converge towards the same density monatomic system
curve as the temperature is increased. At the high temperatures, the
mass difference plays no appreciable role, and the thermal
conductivity only depends on density. At the lowest temperature, 10
K, the thermal conductivity data do not approach the monatomic
value, but do show the density-predicted trend at the large and
small mass ratios. This difference between the low and high
temperature behaviors is due to the difference in the degree of
scattering. At high temperatures, the anharmonic effects are strong
and tend to override the structural effects.

In the cross-plane direction, the data is never close to the
same-density monatomic system, other than at $R_m = 1$. At the two
low temperatures, the data follow the density trend closely, but
with values lower than the monatomic system (this is a structural
effect). At the high temperature, 70 K, there is an increase towards
and decrease away from the $R_m = 1$ thermal conductivity value for
increasing mass ratio. Here, we hypothesize that the effect of the
increasing density is offset by the decrease in phonon scattering in
the cross-plane direction as the two masses in the system approach
the same value.

The general trends in the in-plane and cross-plane directions are
different: in the in-plane direction, higher temperature leads to
the density trend and apparent structure independence, while in the
cross-plane direction, the density trend is found at lower
temperatures where structural effects are important. While this
difference is most likely a result of the anisotropic nature of the
system, the underlying reasons are not yet clear. Qualitatively, a
thermal conductivity following the density trend suggests strong
contributions from phonon modes in which the atoms in the unit cell
move cooperatively -- a characteristic of the acoustic phonon modes.
In general, from these results we see that having a polyatomic
structure introduces additional decreases in thermal conductivity
beyond that predicted by density trends.

\section{\label{S-corr}Correlation between phonon band structure and thermal
conductivity}

In this section, we address the link between thermal conductivity
and phonon dispersion (C and C$^*$ in Fig$.$ \ref{F-3way}), without
explicit consideration of the atomic structure.

\subsection{\label{SS-disp-hcacf}Dispersion and HCACF}

\begin{figure}
\includegraphics{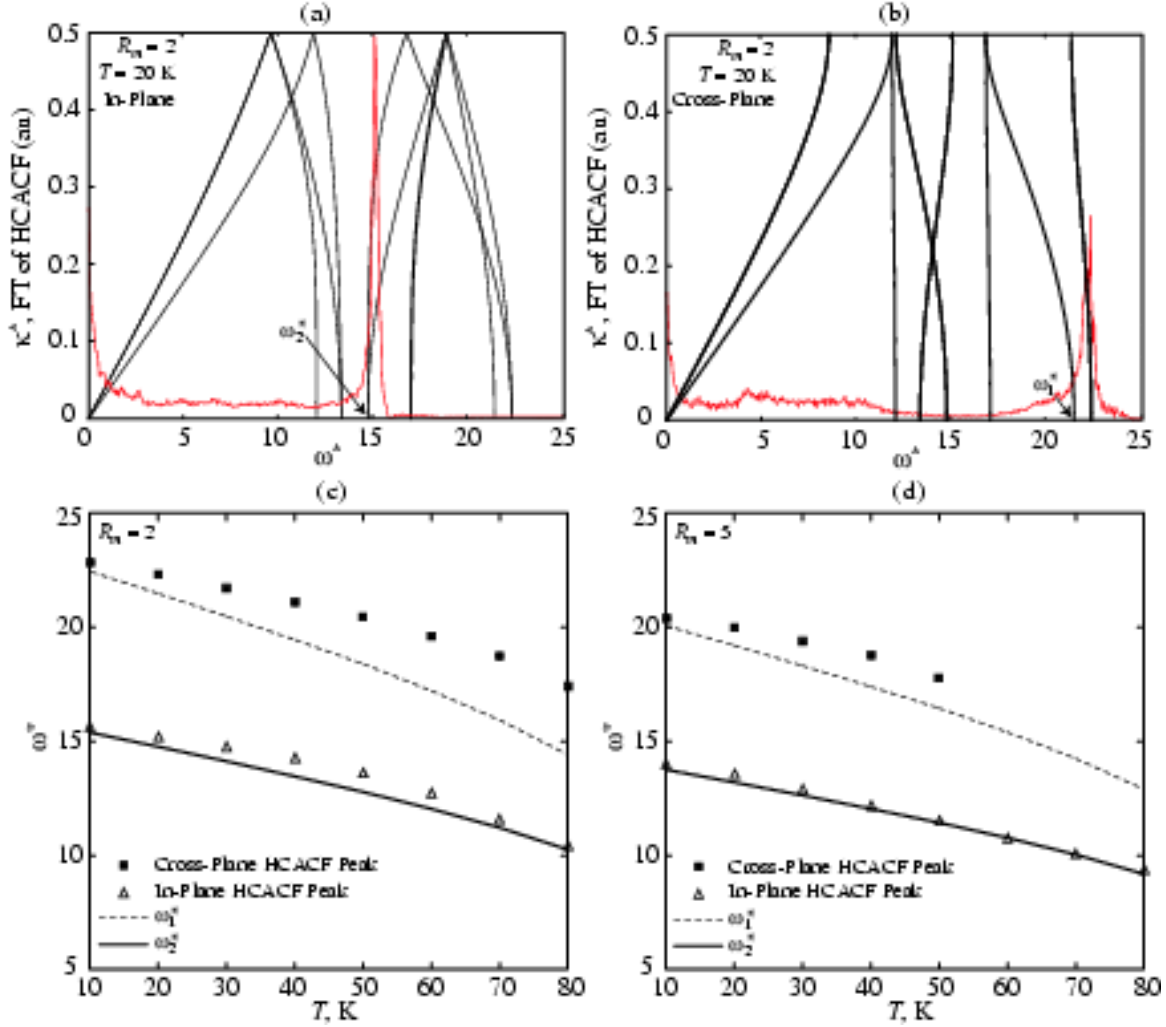}
\caption{\label{F-peaks} (a) In-plane and (b) cross-plane dispersion
curves superimposed on the respective HCACF spectra for $R_m=2$,
$T=20$ K. Comparison of the peaks in the HCACF to the associated
frequencies in the quasi-harmonic dispersion curves for (c) $R_m=2$
and (d) $R_m=5$ in the in-plane and cross-plane directions, plotted
as a function of temperature. (Color online)}
\end{figure}

While peaks in the spectrum of the HCACF have been
observed,\cite{che2000,dong2001,mcgaughey2004b} no comprehensive
explanation has been given for their origin. It has been
hypothesized that they corresponded to optical phonons,
\cite{lindan1991,mcgaughey2004b} due to their frequencies lying in
that region of the phonon spectrum. Such a result is shown in Fig$.$
\ref{F-peaks}(a), where the quasi-harmonic phonon dispersion curves
for the in-plane direction are shown on top of the HCACF spectra for
the in-plane and cross-plane directions for $R_m = 2$ and a
temperature of 20 K. In Fig$.$ \ref{F-peaks}(b), the location of the
peak [as found in the decomposition of the HCACF with Eq$.$
(\ref{E-qdecomp})] is plotted as discrete data points as a function
of temperature for the in-plane and cross-plane directions in the
$R_m=2$ system. The corresponding data for $R_m=5$ are plotted in
Fig$.$ \ref{F-peaks}(c).

By overlaying the dispersion curves and the HCACF spectra for both
directions for a range of temperatures and mass ratios, we find that
the peak in the in-plane spectrum is coincident with one of the
dispersion branches at zero wave-vector (i.e., the gamma point --
the center of the Brillouin zone). The associated quasi-harmonic
frequency is denoted by $\omega_2^*$ in Fig$.$ \ref{F-peaks}(a), and
is plotted as a continuous line in Figs$.$ \ref{F-peaks}(b) and
\ref{F-peaks}(c). A similar coincidence exists for the cross-plane
peak, which is consistently near the gamma point frequency
$\omega_1^*$, also shown in Fig$.$ \ref{F-peaks}(a). These
frequencies are also plotted in Figs$.$ \ref{F-peaks}(b) and
\ref{F-peaks}(c). The link between HCACF peaks and gamma-point
frequencies has also been found in LJ SL structures with longer
period lengths.\cite{landry2006}

The agreement between the HCACF peaks and the dispersion frequencies
is very good, especially for the in-plane peak. The data for
$R_m=10$ shows a similar agreement as for $R_m=5$. The specification
of the peak from the HCACF decomposition becomes difficult at high
temperatures, particularly in the cross-plane direction [this can be
seen in Fig$.$ \ref{F-whcacf}(b)]. Quasi-harmonic theory will lead
to an under-prediction of the dispersion curve frequencies, which
seems to be the case for the cross-plane peak.

From these results, it is clear that specific optical phonon modes
generate the peaks in HCACF. While these modes do not make a
significant contribution to the thermal conductivity, as will be
discussed in Section \ref{SS-disp-k}, we interpret the appearance of
peaks in the HCACF as an indication that they are partly responsible
for the increased phonon scattering (and lower thermal conductivity)
in the diatomic system.

\subsection{\label{SS-disp-k}Dispersion and thermal conductivity}

As seen in Figs$.$ \ref{F-kT}(a) and \ref{F-kT}(b), the thermal
conductivities in the in-plane and cross-plane directions are lower
than the monatomic value. The deviation from the monatomic value
increases as the mass ratio increases. In the limit of an infinite
mass ratio, the thermal conductivity will go to zero. This limit,
and the overall behavior, can be partly understood from the trends
in the dispersion curves. As the mass ratio increases, the
dispersion branches get flatter (see Fig$.$ \ref{F-disp}). The
phonon group velocities (related to the slope of the dispersion
curves) will get correspondingly smaller until they reach a value of
zero when the branches are completely flat, leading to zero thermal
conductivity.

It is often assumed, although in most cases not properly justified,
that acoustic phonons are the primary energy carriers in a
dielectric crystal. Optical phonons are assumed to have a small
contribution because of their supposed lower group velocities. Yet
from a visual inspection of the dispersion curves in Fig$.$
\ref{F-disp}, one can see that there is not a significant difference
in the slopes of the acoustic and optical branches. Based on the
results of the thermal conductivity decomposition by Eq$.$
(\ref{E-kdecomp}) (not presented in detail here), the acoustic modes
are responsible for most of the thermal conductivity. For a mass
ratio of 2 in the cross-plane direction, the acoustic contribution
ranges between 99.98\% at a temperature of 10 K to 99.79\% at a
temperature of 80 K. For a mass ratio of 10 in the cross-plane
direction, the acoustic contribution ranges between 98.84\% at a
temperature of 10 K to 85.08\% at a temperature of 80 K. Thus, the
dispersion trends alone cannot completely explain the observed
behavior, and changes in the phonon scattering behavior must be
considered.

\begin{figure}
\includegraphics{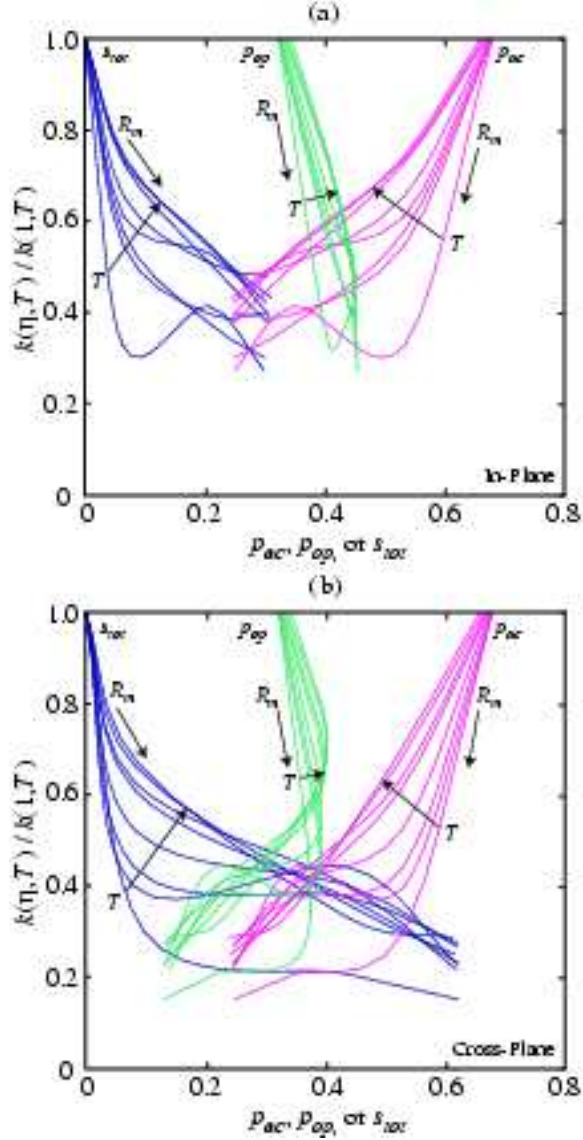}
\caption{\label{F-kbands} Thermal conductivity at all temperatures
and mass ratios considered, scaled by the appropriate $R_m=1$ value,
plotted against the phonon dispersion metrics $p_{ac}$, $p_{op}$,
and $s_{tot}$. To best see the trends, smooth curves through each of
the data sets are plotted. (Color online)}
\end{figure}

In Figs$.$ \ref{F-kbands}(a) and \ref{F-kbands}(b), the thermal
conductivities for all cases considered in this study are plotted as
a function of the dispersion metrics  $p_{ac}$, $p_{op}$, and
$s_{tot}$ introduced in Section \ref{S-dispersion} and plotted in
Fig$.$ \ref{F-disp-gen}(b). These quantities represent the fraction
of the frequency spectrum taken up by the acoustic phonon modes, the
optical phonon modes, and the stop bands. All thermal conductivities
are scaled by the corresponding $R_m=1$ value at the same
temperature.

From these plots, a number of general trends can be discerned.
First, in both the in-plane and cross-plane directions, as the
relative extent of the acoustic phonon modes increases, the thermal
conductivity increases. As discussed, the acoustic modes are most
responsible for the thermal conductivity. A larger frequency extent
will lead to larger acoustic phonon group velocities, and a higher
thermal conductivity. As the extent of the stop bands increases, the
thermal conductivity decreases in both directions. In a stop band,
there is no phonon wave propagation and hence no means for thermal
transport. The optical phonon effect is different in the two
directions, consistent with the behavior of $p_{op}$ seen in Fig$.$
\ref{F-disp-gen}(b). Initially, in both directions, an increase in
$p_{op}$ leads to a decrease in the thermal conductivity. The effect
of the optical branches is similar to the effect of the stop bands:
an accompanying decrease in the extent of the acoustic phonon modes.
This trend continues for the in-plane direction, but the cross-plane
direction shows a different behavior, with a decrease in $p_{op}$ as
the thermal conductivity decreases. This change of behavior
motivates further research on the role of the optical modes, but at
least indicates that considering only one of $p_{ac}$, $p_{op}$, or
$s_{tot}$ is not sufficient, and that all must be considered to
completely understand the behavior of a given system.

These results can also shed light on the competition between
dispersion characteristics  and anharmonic scattering in affecting
thermal conductivity. In a harmonic system energy transport can be
completely described by the dispersion curves, as there are no mode
interactions. As anharmonicity is introduced to such a system (e.g.,
by increasing the temperature), the dispersion curves will no longer
be sufficient to predict the transport as the modes will interact.
From Figs$.$ \ref{F-kbands}(a) and \ref{F-kbands}(b), we see that as
the temperature increases, the thermal conductivity is less
sensitive to changes in the dispersion metrics (i.e., the magnitude
of the slopes of the curves decrease), consistent with the above
discussion

The analysis in this study suggests that a spectral perspective to
thermal transport, quantified by dispersion metrics such as those
proposed, provides the following benefits: (i) it enables a
quantitative analysis (e.g., by evaluating the slopes of the thermal
conductivity versus dispersion metrics curves) of the relative
contributions of dispersion and anharmonic scattering mechanisms to
thermal transport, and (ii) it uncouples the structure-property
relation (as discussed in Section \ref{S-k}) to that of
structure-dispersion and dispersion-property relations, providing a
simplified analysis framework that could be taken advantage of in
design studies.

\section{Conclusions \label{S-summary}}

We have studied a layered diatomic Lennard-Jones crystal with
varying mass ratio in order to establish a temperature-dependent
three-way correlation between (i) the unit cell structure and
inter-atomic potential, (ii) the associated phonon dispersion
curves, and (iii) the bulk thermal conductivity (see Fig$.$
\ref{F-3way}). The dispersion curves were obtained using lattice
dynamics calculations, and their characteristics were quantified
using a set of novel band diagram metrics. The thermal conductivity
was predicted using molecular dynamics simulations and the
Green-Kubo method.

As shown in Fig$.$ \ref{F-kT}, the thermal conductivity of the
layered material is less than that of a corresponding monatomic
system, and is directionally dependent. To understand the thermal
conductivity trends, one must consider the effects of density and
crystal structure, with consideration to both the dispersion and the
anharmonic scattering characteristics (see Fig$.$ \ref{F-km}).

The HCACF for each material was fit to an algebraic function,
allowing for the roles of acoustic and optical phonons to be
distinguished. The peaks in the FT of the HCACF were found to
coincide with frequencies obtained from the dispersion calculations,
confirming that these peaks are generated by optical phonons (see
Fig$.$ \ref{F-peaks}).

The dispersion curves metrics introduced in Section
\ref{S-dispersion}, and plotted in Fig$.$ \ref{F-disp-gen}, describe
the frequency extent of the acoustic and optical phonon modes and
that of the stop bands. As shown in Fig$.$ \ref{F-kbands}, the
thermal conductivity increases with increasing $p_{ac}$, and
decreases with increasing $s_{tot}$. The dependence on $p_{op}$
varies depending on the mass ratio and the crystallographic
direction, a behavior that motivates further investigation.

The rate of change of the thermal conductivity versus the dispersion
metrics (see Fig$.$ \ref{F-kbands}) decreases with increasing
temperature, consistent with the increasing role of anharmonic
scattering on thermal transport at higher temperature. These
relations provide an indirect measure of the relative contributions
of dispersion and anharmonic scattering to thermal transport.
Moreover, the dispersion metrics allow for the uncoupling of the
traditional thermal conductivity structure-property relation to that
of structure-dispersion and dispersion-property relations. This
decomposition facilitates further elucidation of the underlying
physics of thermal transport, and could be used as an effective tool
for material design.

\begin{acknowledgments}
This work was partially supported by the U.S. Department of Energy,
Office of Basic Energy Sciences under grant DE-FG02-00ER45851 (AJHM,
MK), and the Horace H. Rackham School of Graduate Studies at the
University of Michigan (AJHM). Some of the computer simulations were
performed using the facilities of S. R. Phillpot and S. B. Sinnott
in the Materials Science and Engineering Department at the
University of Florida.
\end{acknowledgments}

\appendix*
\section{Lattice Dynamics Calculations \label{S-appendix}}

In this appendix, we outline the analytical procedure used to
investigate the lattice dynamics. The formulation is based on that
presented by Dove\cite{dove1993} and Maradudin.\cite{maradudin1974}

Determining phonon mode frequencies and polarization vectors using
lattice dynamics is greatly simplified under the harmonic
approximation. In this approximation, a Taylor series expansion of
the system potential energy about its minimum value is truncated
after the second-order term. This approach leads to harmonic
equations of motion that can be solved analytically for simple
systems,\cite{kittel1996} but are most easily solved numerically.

Given an interatomic potential and the equilibrium positions of the
atoms in a crystal with an $n$-atom unit cell, the analysis proceeds
as follows:

The equation of motion of the $j$-th atom ($j = 1, 2, ..., n$) in
the $l$-th unit cell is
\begin{equation}
m_j {\bf{\ddot{u}}} (jl,t) =
\sum_{j'l'}{\bf{\Phi}}\left(jl;j'l'\right)\cdot{\bf{u}}(jl,t),
\label{A-EOM}
\end{equation}
where $m_j$ is the mass of atom $j$, $\bf{\Phi}$ is the $3 \times 3$
force constant matrix describing the interaction between atoms $jl$
and $j'l'$, $\mathbf{u}(jl,t)$ is the displacement of atom $jl$ from
its equilibrium position, and the summation is over every atom in
the system (including $jl$ itself). The elements of $\bf{\Phi}$ are
given by
\begin{equation}
\mathbf{\Phi}_{\alpha\beta}(jl;j'l')=\left\{ \begin{array}{ll}
-\chi_{\alpha\beta}(jl;j'l') & jl\neq j'l' \\
\\
\displaystyle\sum_{j''l''} \chi_{\alpha\beta}(jl;j''l'') & jl =
j'l',
\end{array} \right. \label{A-Phi}
\end{equation}
where $\alpha$ and $\beta$ can be the cartesian coordinates $x$,
$y$, and $z$. The summation in Eq$.$ (\ref{A-Phi}) is over pairings
between atom $jl$ and every atom in the system other than itself.
For a two-body potential, $\phi(r)$, $\chi_{\alpha\beta}$ is
\begin{eqnarray}
\chi_{\alpha\beta}(jl;j'l')&=&\chi_{\alpha\beta}(\pmb{r}_{jl}-\pmb{r}_{j'l'})
=\left.\frac{\partial^2\phi(r)}{\partial r_\alpha
\partial r_\beta}\right|_{r=|\pmb{r}_{jl}-\pmb{r}_{j'l'}|}\nonumber\\
&=&\left.\left\{ \frac{r_\alpha r_\beta}{r^2}\left[\phi^{\prime
\prime}(r) - \frac{1}{r}\phi^\prime(r) \right] +
\frac{\delta_{\alpha\beta}}{r}\phi^\prime(r)
\right\}\right|_{r=|\pmb{r}_{jl}-\pmb{r}_{j'l'}|},
\end{eqnarray}
where $^\prime$ indicates a derivative with respect to $r$,
$\delta_{\alpha\beta}$ is the delta function, $\pmb{r}$ is a
position vector, and $r_{\alpha}$ is the $\alpha$- component of
$\pmb{r}$.

The solution to Eq$.$ (\ref{A-EOM}) is assumed to be harmonic and is
a summation over the normal (phonon) modes of the system, which are
traveling waves. Each normal mode has a different wave vector
$\pmb{\kappa}$ and dispersion branch $\nu$ such that
\begin{equation}
{\bf{u}}(jl,t)=\sum_{\pmb{\kappa},
\nu}m_j^{-1/2}{\bf{e}}_j(\pmb{\kappa},\nu)\exp\{i[\pmb{\kappa} \cdot
{\bf{r}}(jl) - \omega(\pmb{\kappa}, \nu)t]\}, \label{A-solution}
\end{equation}
where $\omega$ is the mode frequency and $t$ is time. The allowed
wave vectors are set by the crystal structure. Substituting Eq.
(\ref{A-solution}) into Eq. (\ref{A-EOM}) leads to the eigenvalue
equation
\begin{equation}
\omega^2(\pmb{\kappa},\nu){\bf{e}}(\pmb{\kappa},\nu)={\bf{D}}(\pmb{\kappa})
\cdot {\bf{e}}(\pmb{\kappa},\nu) \label{A-eigenvalue},
\end{equation}
where $\bf{D}(\pmb{\kappa})$ is the $3n \times 3n$ dynamical matrix
and $\bf{e}(\pmb{\kappa}, \nu)$ is the polarization vector of length
$3n$. In Eq. (\ref{A-solution}), ${\bf{e}}_j(\pmb{\kappa}, \nu)$
contains the $3$ elements of $\bf{e}(\pmb{\kappa}, \nu)$ that
correspond to the $j$-th atom of the unit cell (elements $3j-2$,
$3j-1$, and $3j$). The elements of $\bf{D}(\pmb{\kappa})$ are given
by
\begin{equation}
D_{3(j-1)+\alpha, 3(j'-1)+\beta}(jj',
\pmb{\kappa})=\frac{1}{(m_jm_{j'})^{1/2}}\sum_{l'}{\bf{\Phi}}_{\alpha\beta}(j0;
j'l')\exp(i\pmb{\kappa} \cdot[{\bf{r}}(j'l')-{\bf{r}}(j0)]).
\label{A-D}
\end{equation}
Here, the summation is over all unit cells and ($alpha$ or $\beta$
equals 1 for $x$, etc.). For example, in a unit cell of four atoms
denoted by A, B, C and D (like that shown in Fig$.$
\ref{F-structure}), the summation in Eq$.$ (\ref{A-D}) for
($j=1,j'=2$) would be over pairings between atom A in the 0-th unit
cell and atom B in all unit cells, including 0.

The dynamical matrix is Hermitian and therefore has real eigenvalues
and orthogonal eigenvectors. The square roots of the eigenvalues of
$\bf{D}(\pmb{\kappa})$ are the phonon mode frequencies
$\omega(\pmb{\kappa},\nu)$, and the eigenvectors of
$\bf{D}(\pmb{\kappa})$ give the phonon polarization vectors
$\mathbf{e}(\pmb{\kappa},\nu)$. The polarization vectors (also known
as the mode shapes) are normalized such that
\begin{equation}
[{\bf{e}}(\pmb{\kappa},\nu)]^T \cdot
[{\bf{e}}(\pmb{\kappa},\nu)]^*=1.
\end{equation}
In general, the elements of $\mathbf{e}(\pmb{\kappa},\nu)$ are
complex and describe the relative amplitude and phase difference of
the displacements of the atoms in the unit cell.

Dispersion curves can be generated by calculating the phonon mode
frequencies for a range of wave vectors in a particular direction.
For example, the dispersion curve presented in Fig$.$
\ref{F-disp}(b) was generated using 100 evenly spaced wave vectors
in the in-plane direction with wave numbers ranging from 0 to
$\pi/a$ where $a$ is the lattice constant. The dispersion curve has
3 acoustic branches and $3n-3$ optical branches. The dispersion
shown in Fig$.$ \ref{F-disp}(c) has only 8 branches because the
transverse branches in this direction are degenerate.

The procedure outlined here gives results that are exact at zero
temperature. For higher temperatures, the zero-pressure lattice
positions should be used (i.e. the quasi-harmonic
approximation).\cite{dove1993} The anharmonic contributions to the
lattice dynamics analysis can be included using molecular dynamics
simulations.\cite{mcgaughey2004c}


\begin{thebibliography}{35}
\bibitem[{sca()}]{scatter}
\bibinfo{note}{In this study, where the phonon transport is diffusive, we will
  use the term scattering to indicate multi-phonon interactions brought about
  by the anharmonic effects present at finite temperatures.}
\end{thebibliography}

\end{document}